\documentclass{FM6_Focus_in_Astronomy}
\usepackage{graphicx}
\usepackage{hyperref}
\usepackage{bm}

\title[FM6.~~Angular momentum from surveys]
{Emerging angular momentum \\ physics from kinematic surveys}

\author[Matthew Colless]{Matthew Colless$^1$}

\affiliation{$^1$Research School of Astronomy and Astrophysics,
                         Australian National University \\
                         Canberra, ACT 2611, Australia \\
                         email: {\tt matthew.colless@anu.edu.au}}

\pubyear{2018}
\jname{Galactic Angular Momentum} 
\editors{Danail Obreschkow, ed.}

\begin{document}

\maketitle

\begin{abstract}
  I review the insights emerging from recent large kinematic surveys of
  galaxies at low redshift, with particular reference to the SAMI,
  CALIFA and MaNGA surveys.  These new observations provide a more
  comprehensive picture of the angular momentum properties of galaxies
  over wide ranges in mass, morphology and environment in the
  present-day universe. I focus on the distribution of angular momentum
  within galaxies of various types and the relationship between mass,
  morphology and specific angular momentum. I discuss the implications
  of the new results for models of galaxy assembly.
\end{abstract}

\firstsection 
\section{Introduction}

This brief review focusses on recent integral field spectroscopy surveys
of the stellar kinematics in large samples of galaxies at low
redshifts. It does not cover radio HI surveys of the neutral gas in
low-redshift galaxies (which are important for understanding the
kinematics at large radius) nor does it extend to surveys at high
redshifts (which explore the origin and evolution of galaxy kinematics).
What {\em local} surveys of stellar kinematics can tell us about angular
momentum in galaxies is its dependence on mass, morphology and other
properties (if sample selection is understood) and its dependence on
environment (if the sample is embedded in a fairly complete redshift
survey); such dependencies can provide {\em indirect} evidence for the
origin and evolution of angular momentum.

It is immediately apparent that all current kinematic surveys have
weaknesses relating to the trade-offs demanded by instrumental
constraints: firstly, between spatial resolution and spatial coverage
(also between spectral resolution and spectral coverage) and, secondly,
between this per-galaxy information and sample size (also sample volume
and completeness). The lack of radial coverage is a serious problem for
late-type disk galaxies having exponential mass profiles (i.e.\ having
Sersic index $n \approx 1$), for which $M/M_{\rm tot}=0.5$,0.8 at
$R/R_e \approx 1.0$,1.8 and $j/j_{\rm tot} = 0.5$,0.8 at
$R/R_e \approx 1.0$,2.2. But it is a much worse problem for early-type
spheroidal galaxies with deVaucouleurs profiles ($n \approx 4$), for
which $M/M_{\rm tot}=0.5$,0.8 at $R/R_e \approx 1.0$,3.2 and
$j/j_{\rm tot} = 0.5$,0.8 at $R/R_e \approx 4.4$,$>$9 (see
Figure~1a). This problem is compounded by the necessary instrumental
trade-off between radial coverage (field of view) and spatial resolution
(spaxel scale) of integral field units (IFUs) due to constraints imposed
by the limited available detector area. For example, in the SAMI sample
the median major axis is $R_e=4.4$\,arcsec (10–-90\% range spans
1.8-–9.4\,arcsec) which means that the SAMI IFUs only sample out to a
median radius of 1.7$R_e$ (see Figure~1b).

\begin{figure}
\begin{center}
  \includegraphics[width=0.4\textwidth]{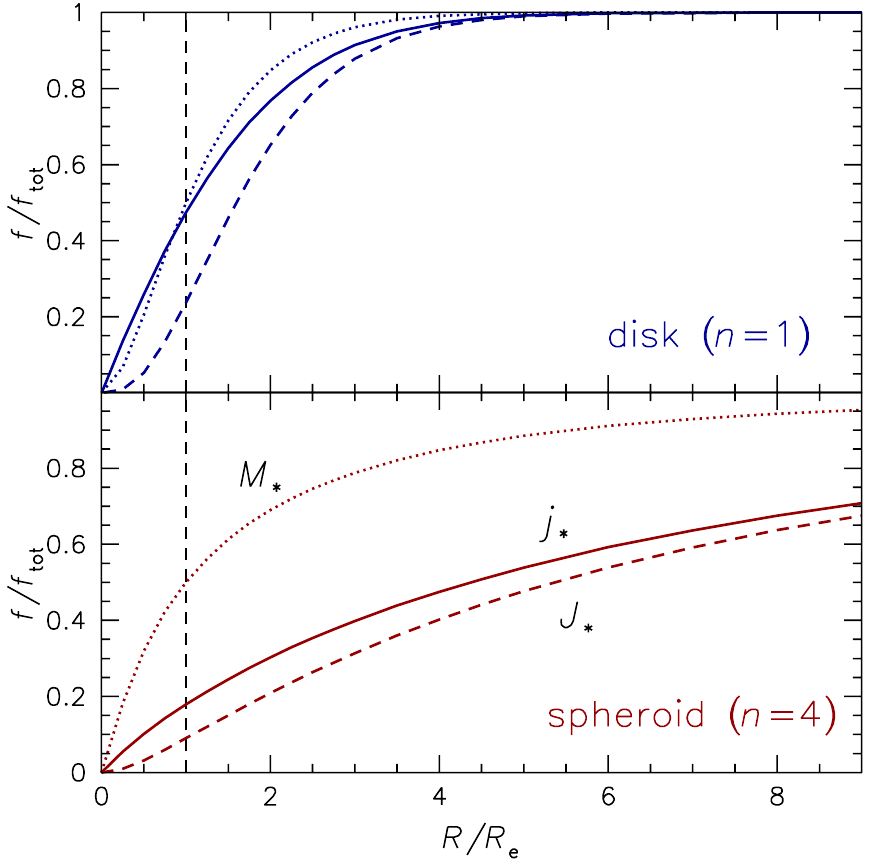} ~~~~~
  \includegraphics[width=0.55\textwidth]{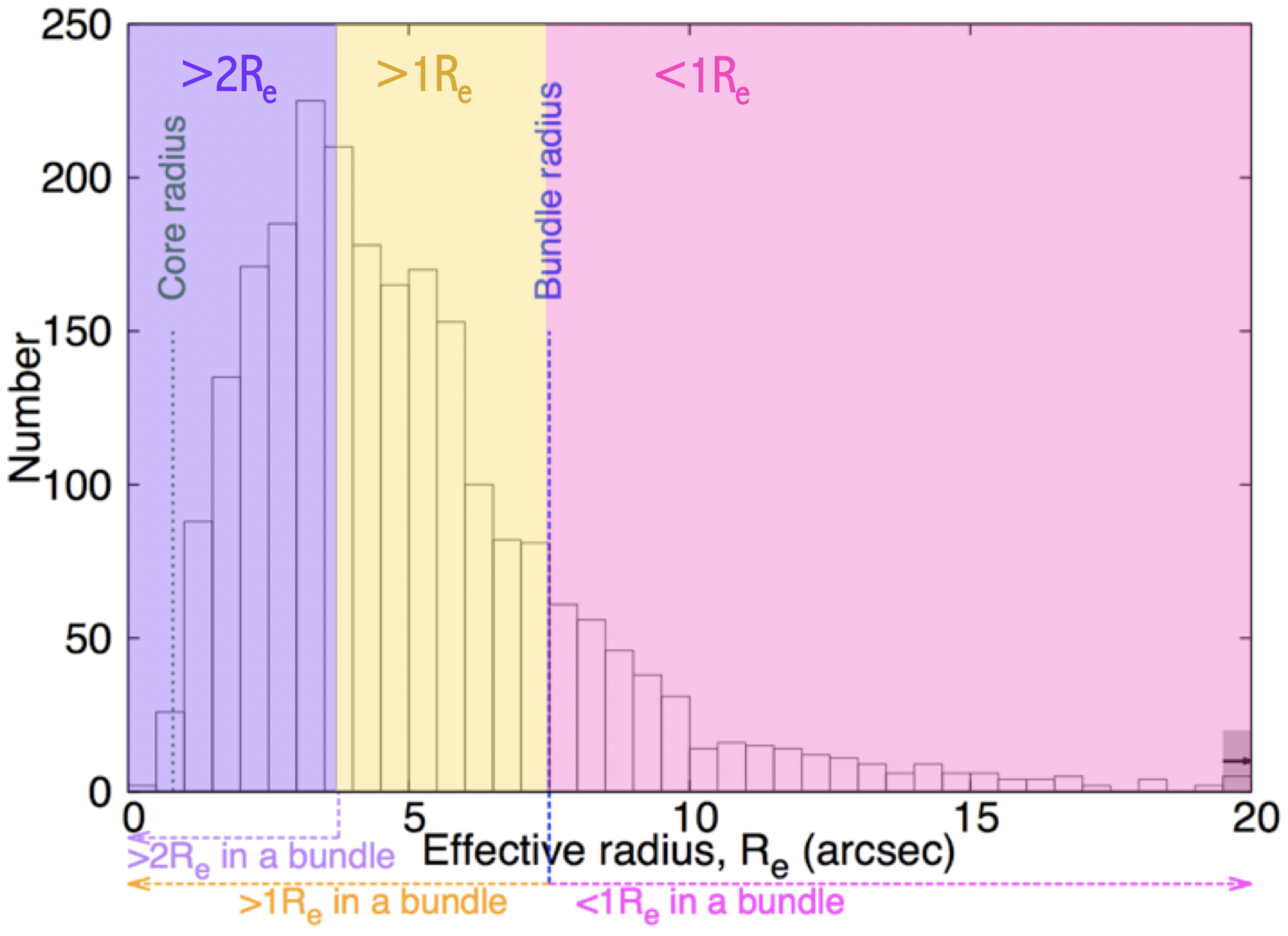}
  \caption{(a)~Left panel: the fraction of mass ($M$), angular momentum
    ($J$), and specific angular momentum ($j=J/M$) as functions of
    radius (in units of the effective radius, $R_e$) for both an
    exponential disk profile (Sersic index $n=1$; top panel) and a
    deVaucouleurs spheroid profile (Sersic index $n=4$; bottom panel)
    [\cite{Romanowsky+Fall2012}, Fig.3]. (b)~Right panel: the
    distribution of effective radius $R_e$ (in arcsec) for the SAMI
    galaxy sample, showing those parts of the sample for which the
    integral field unit covers $<$1$R_e$, $>$1$R_e$ and $>$2$R_e$
    [based on \cite{Green+2015}, Fig.1].}
   \label{fig1}
\end{center}
\end{figure} 

\section{Surveys}

\subsection{SAMI}

SAMI is the Sydney-AAO Multi-IFU instrument on the 3.9m Anglo-Australian
Telescope (AAT). It has 13 IFUs that can be positioned over a 1\,degree
field at the telescope's prime focus. Each hexabundle IFU has
61\,$\times$\,1.6\,arcsec fibres covering a 15\,arcsec diameter field of
view. SAMI feeds the AAOmega spectrograph, which gives spectra over
375--575nm at $R$\,$\approx$\,1800 (70\,km\,s$^{-1}$) and 630--740nm at
$R$\,$\approx$\,4300 (30\,km\,s$^{-1}$). The SAMI Second Data Release
(DR2) includes 1559 galaxies (about half the full sample) covering
0.004\,$<$\,z\,$<$\,0.113 and
7.5\,$<$\,$\log(M_*/M_\odot)$\,$<$\,11.6. The core data products for
each galaxy are two primary spectral cubes (blue and red), three
spatially binned spectral cubes, and a set of standardised aperture
spectra.  For each core data product there are a set of value-added data
products, including aperture and resolved stellar kinematics, aperture
emission line properties, and Lick indices and stellar population
parameters.  The data release is available online through AAO Data
Central (\url{datacentral.org.au}).

\subsection{CALIFA}

CALIFA is the Calar Alto Legacy Integral Field survey, consisting of
integral field spectroscopy for 667 galaxies obtained with PMAS/PPak on
the Calar Alto 3.5m telescope. There are three different spectral
setups: 375--750\,nm at 0.6\,nm FWHM resolution for 646 galaxies,
365--484\,nm at 0.23\,nm FWHM resolution for 484 galaxies, and a
combination of these over 370--750\,nm at 0.6\,nm FWHM resolution for
446 galaxies. The CALIFA Main Sample spans 0.005\,$<$\,$z$\,$<$\,0.03
and the colour-magnitude diagram, with a wide range of stellar masses,
ionization conditions and morphological types; the CALIFA Extension
Sample includes rare types of galaxies that are scarce or absent in the
Main Sample.

\begin{figure}
\begin{center}
  \includegraphics[width=\textwidth]{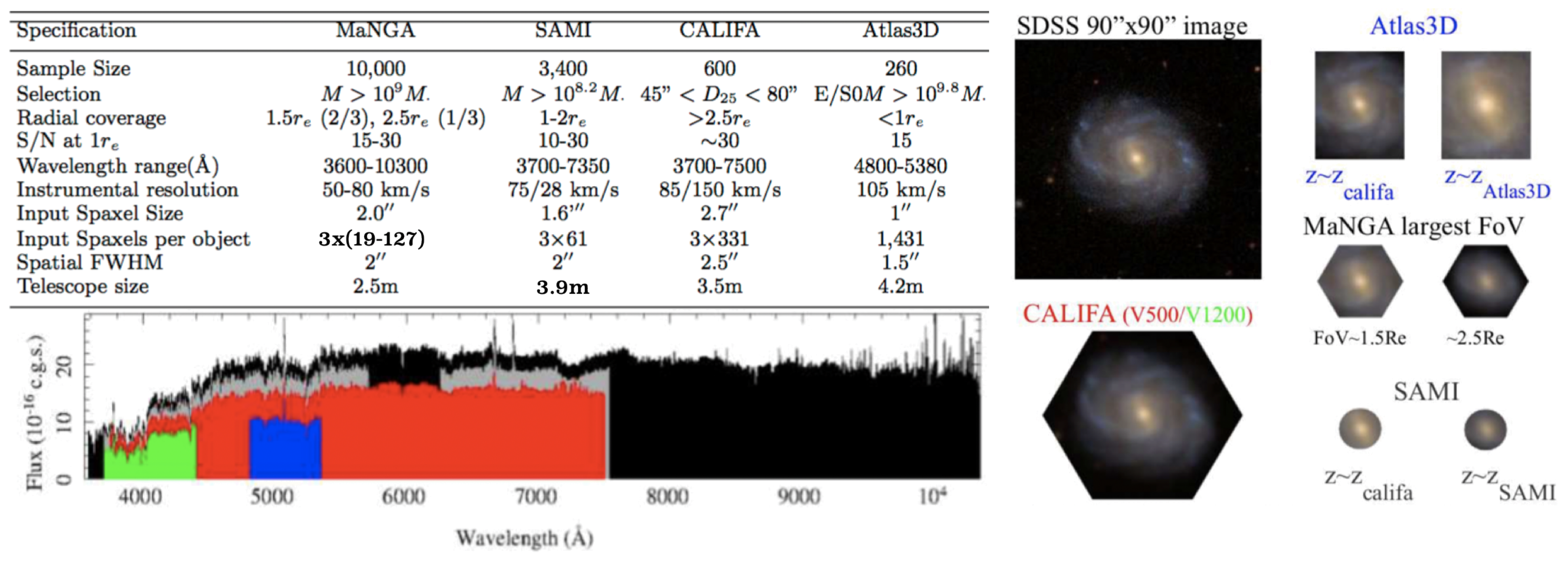}
  \caption{Upper left: table of key parameters of the MaNGA, SAMI,
    CALIFA and Atlas3D surveys. Lower left: the wavelength coverage of
    the MaNGA (black), SAMI (grey), CALIFA (red \& green) and Atlas3D
    (blue) surveys. Right: illustration of the relative fields of view
    covered by the IFUs used in each survey. [Based on
    \cite{Sanchez+2015}, Table~1 \& Figure~1.]}
   \label{fig2}
\end{center}
\end{figure} 

\subsection{MaNGA}

MaNGA is the Mapping Nearby Galaxies at Apache Point Observatory survey
(part of SDSS-IV). It is studying the internal kinematic structure and
composition of gas and stars in 10,000 nearby galaxies. It employs 17
fibre-bundle IFUs varying in diameter from 12\,arcsec (19 fibres) to
32\,arcsec (127 fibers) that feed two dual-channel spectrographs
covering 360--1030\,nm at $R$\,$\approx$\,2000. The targets have
$M_*$\,$>$\,$10^9$\,$M_\odot$ based on SDSS-I redshifts and $i$-band
luminosities. The MaNGA sample is designed to approximate uniform radial
coverage in terms of $R_e$, a flat stellar mass distribution, and a
wide range of environments. SDSS Data Release 14 (DR14) includes MaNGA
data cubes for 2812 galaxies.

\subsection{Comparison}

Figure~2 provides a tabular and graphical summary of the parameters of
these three surveys (and also the earlier Atlas3D survey), which helps
to understand their various relative strengths and weaknesses, and
consequently their complementarities. A few kinematic surveys of small
samples offer greater radial coverage and higher velocity resolution:
SLUGGS surveyed kinematics of 25 early-type galaxies to $\sim$3$R_e$
from stars and to $\sim$10$R_e$ using globular clusters (\cite[Bellstedt
\etal\ 2018]{Bellstedt+2018}); PN.S surveyed the kinematics of 33
early-type galaxies to $\sim$10$R_e$ using planetary nebulae
(\cite[Pulsoni \etal\ 2018]{Pulsoni+2018}).

\section{Results}

\subsection{Role of angular momentum}

After mass, angular momentum is the most important driver of galaxy
properties, with a key role in the formation of structure and
morphology. For regular oblate rotators, angular momentum can be derived
from dynamical models as well as direct estimates of projected angular
momentum. Surveys can determine population variations in the total
angular momentum and its distribution with radius, exploring
dependencies on mass, morphology, ellipticity and other
properties. These relations can provide insights on the assembly
histories of galaxies for comparison with simulations.

\subsection{Angular momentum \& spin profiles}

SAMI, CALIFA and MaNGA together now provide angular momentum profiles
(or, alternatively, spin proxy, $\lambda_R$, as a function of $R/R_e$)
for thousands of galaxies to $R/R_e$\,$\sim$\,1 and for hundreds of
galaxies to $R/R_e$\,$\sim$\,2. These samples are large enough to be
useful when split by mass, morphology or environment. Figure~3 shows
spin profiles for galaxies from the CALIFA survey
(\cite[Falc\'on-Barroso 2016]{Falcon-Barroso2016}) and the MaNGA survey
(\cite[Greene \etal\ 2018]{Greene+2018}); \cite{Foster+2018} give
similar results from the SAMI survey.

\begin{figure}
\begin{center}
  \includegraphics[width=0.4\textwidth]{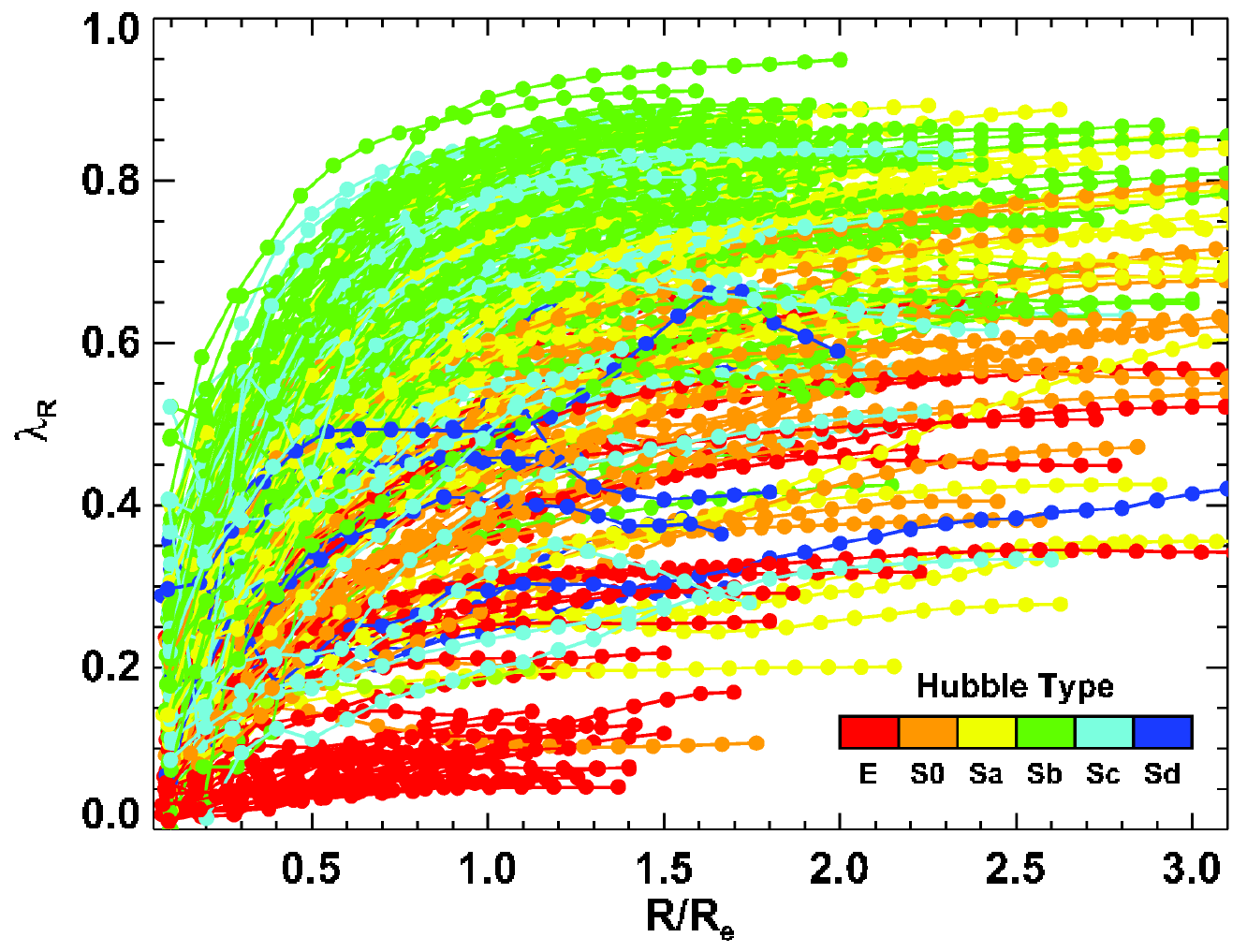} 
  \includegraphics[width=0.59\textwidth]{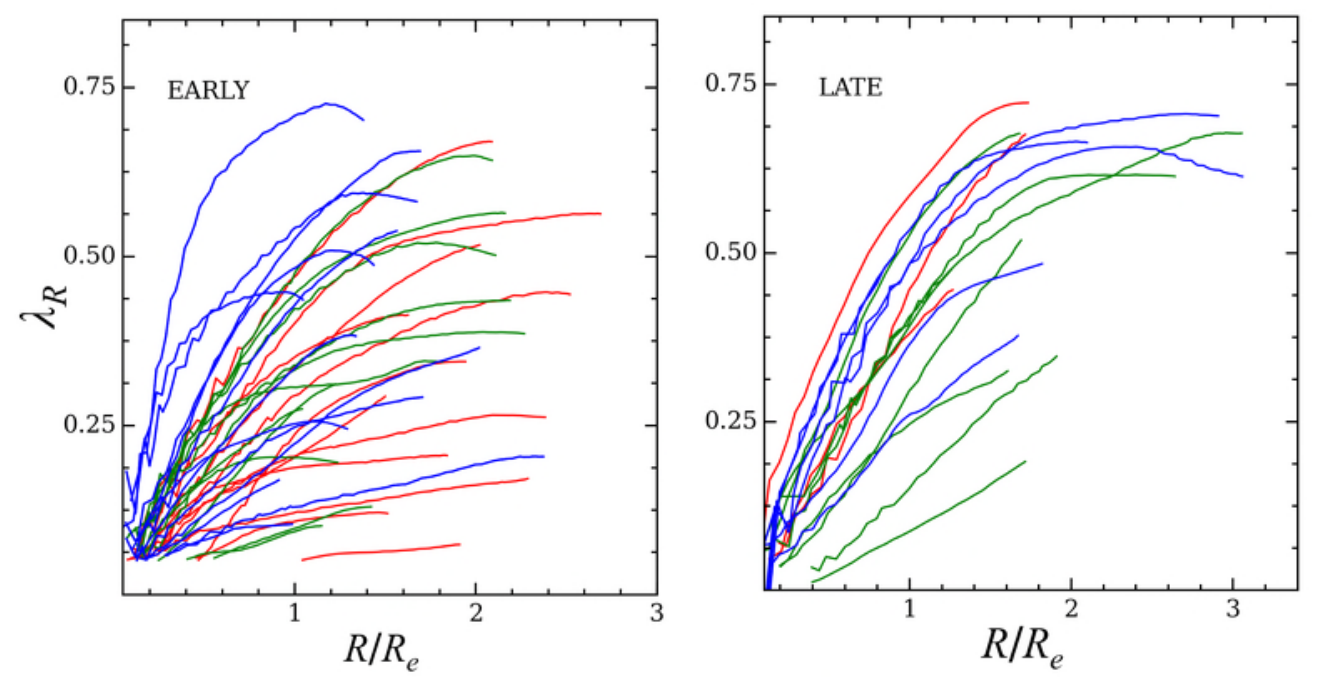}
    \caption{Left: Galaxy spin profiles from CALIFA, showing the variation
    with Hubble type [\cite{Falcon-Barroso2016}, Fig.2]. Right: Galaxy
    spin profiles from MaNGA, showing the variation with mass for early
    and late-type galaxies [\cite{Greene+2018}, Fig.3].}
   \label{fig3}
\end{center}
\end{figure} 

\subsection{Spin, morphology \& ellipticity}

Typical galaxies lie on a plane relating mass $M$, $j$ and stellar
distribution (quantified by, e.g., Sersic index $n$ or photometric
concentration index), with overall morphologies regulated by their mass
and dynamical state (see, e.g., \cite[Cortese \etal\
2016]{Cortese+2016}). The correlation shown in the left panel of
Figure~4 between the offset from the mass--angular momentum ($M$-–$j$)
relation and spin parameter $\lambda_R$ shows that at fixed $M$ the
contribution of ordered motions to dynamical support varies by more than
a factor of three. The right panel of Figure~4 shows that $\lambda_R$
correlates strongly with morphology and concentration index (especially
if slow-rotators are removed), suggesting that late-type galaxies and
early-type fast-rotators form a continuous class in terms of their
kinematic properties.

\begin{figure}
\begin{center}
  \includegraphics[width=0.52\textwidth]{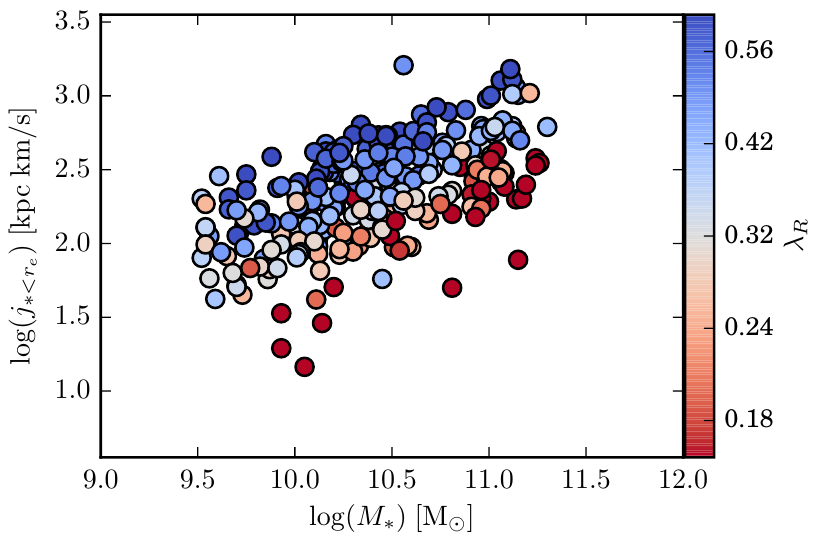}
  \includegraphics[width=0.46\textwidth]{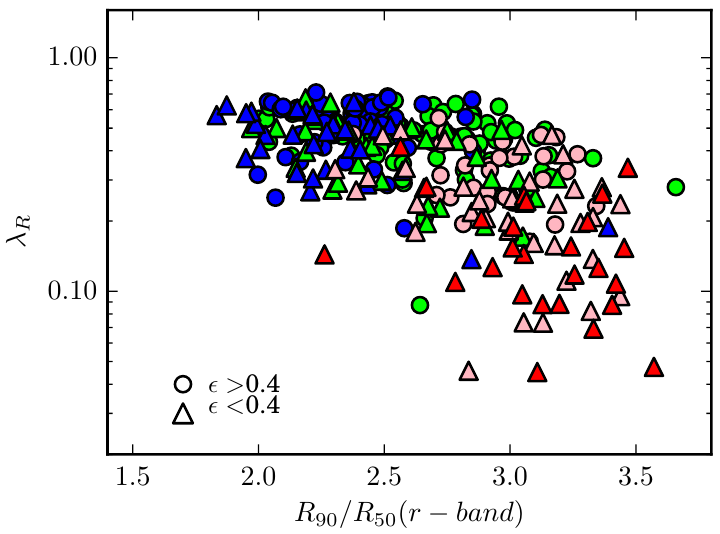}
  \caption{Left: Specific angular momentum versus stellar mass for SAMI
    galaxies, colour-coded by spin parameter $\lambda_R$. Right: Galaxy
    spin parameter versus $r$-band concentration for SAMI galaxies,
    colour-coded by morphology. [Based on \cite{Cortese+2016}, Figs~6
    \&~7].}
   \label{fig4}
\end{center}
\end{figure} 

The spin--ellipticity ($\lambda_R$--$\epsilon$) diagram is a
particularly revealing frame for understanding relations between
kinematic and morphological properties of galaxies. This is illustrated
in Figure~5, from the work of \cite{Graham+2018} using the MaNGA survey.
The left panel shows the strong correlation between the mass of a galaxy
and its position in this diagram, with more massive galaxies tending to
have lower spin and ellipticity. The central panel shows the areas of
the diagram occupied by various morphological types: elliptical galaxies
occupy the low-$\lambda_R$, low-$\epsilon$ region, while lenticular and
spiral galaxies largely overlap, covering the full range of $\epsilon$
at $\lambda_R$\,$>$\,0.5. The right panel shows how galaxies belonging
to different kinematic classes are distributed: spirals generally lie in
the region consistent with rotationally-dominated kinematics, while
regular (fast-rotating) early-type galaxies occupy a wider range of
$\lambda_R$ at given $\epsilon$, with lower $\lambda_R$ corresponding to
systems with more pressure-support; slowly-rotating (`non-rotating')
early-type galaxies mainly occupy the region with $\lambda_R$\,$<$\,0.15
and 0\,$<$\,$\epsilon$\,$<$\,0.4.

\begin{figure}
\begin{center}
  \includegraphics[width=\textwidth]{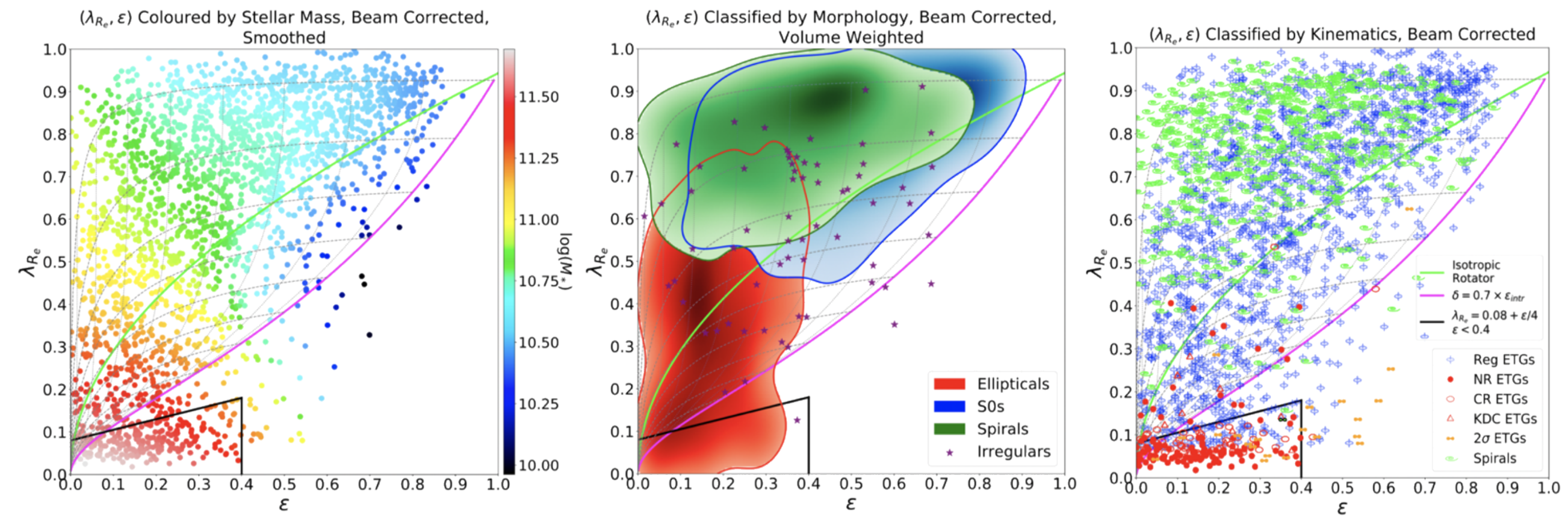}
  \caption{Distributions of galaxy properties in the spin--ellipticity
    ($\lambda_R$--$\epsilon$) diagram: left---stellar mass;
    centre---visual morphology; right---kinematic class.
    [\cite{Graham+2018}, Figs~5,~8~\&~9.]}
   \label{fig5}
\end{center}
\end{figure}

\begin{figure}
\begin{center}
  \includegraphics[width=\textwidth]{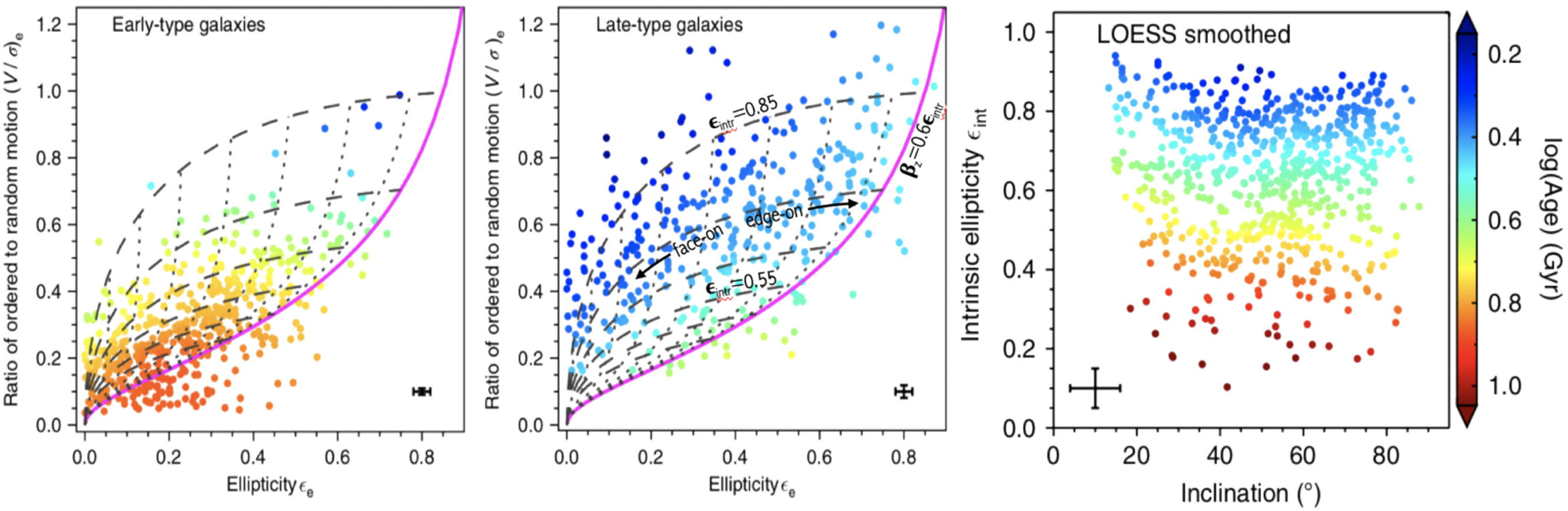}
  \caption{Left/middle panels: The ratio of ordered to random motions
    ($V/\sigma$) versus the apparent ellipticity for early/late-type
    galaxies. Right panel: assuming galaxies are oblate rotators, the
    derifed distribution of intrinsic ellipticity as a function of
    apparent inclination. [\cite{vandeSande+2018}, Figs~3~\&~4.]}
   \label{fig6}
\end{center}
\end{figure} 

There is also an strong correlation between a galaxy's spin parameter
and its intrinsic ellipticity, as demonstrated using the SAMI survey by
\cite{vandeSande+2018}. Figure~6 shows the distribution of the ratio of
rotation velocity to velocity dispersion ($V/\sigma$) with apparent
ellipticity ($\epsilon$) for early-type and late-type galaxies, together
with the inferred distribution of intrinsic ellipticity
($\epsilon_{\rm int}$). This is derived using the theoretical model
predictions for rotating, oblate, axisymmetric spheroids with varying
intrinsic shape and anisotropy shown by the dashed and dotted lines in
the left two panels. The galaxies are colour-coded by the
luminosity-weighted age of their stellar populations, and the righthand
panel shows the clear trend of age with intrinsic ellipticity. As
\cite{vandeSande+2018} discuss in detail, this newly discovered relation
extends beyond the general notion that ‘disks are young’ and ‘bulges are
old’.

\subsection{The mass--angular momentum relation}

The mass--angular momentum ($M$-–$j$) relation is discussed in detail
elsewhere in these proceedings. However, it is worth noting the
opportunties for studying this key relation that follow from large
surveys providing kinematics for many galaxies. Some prelminary results
from the SAMI survey are shown in Figure~7 (D'Eugenio \etal, in prep.),
using hundreds of galaxies with masses and angular momenta derived from
self-consistent dynamical models---in this case, Jeans anisotropic mass
(JAM) models. This permits the study of the $M$-–$j$ relation for
subsets of the population, such as different morphological types. While
the results shown here are too preliminary to allow conclusions to be
drawn, the opportunities are clear.

\begin{figure}
\begin{center}
  \includegraphics[width=0.56\textwidth]{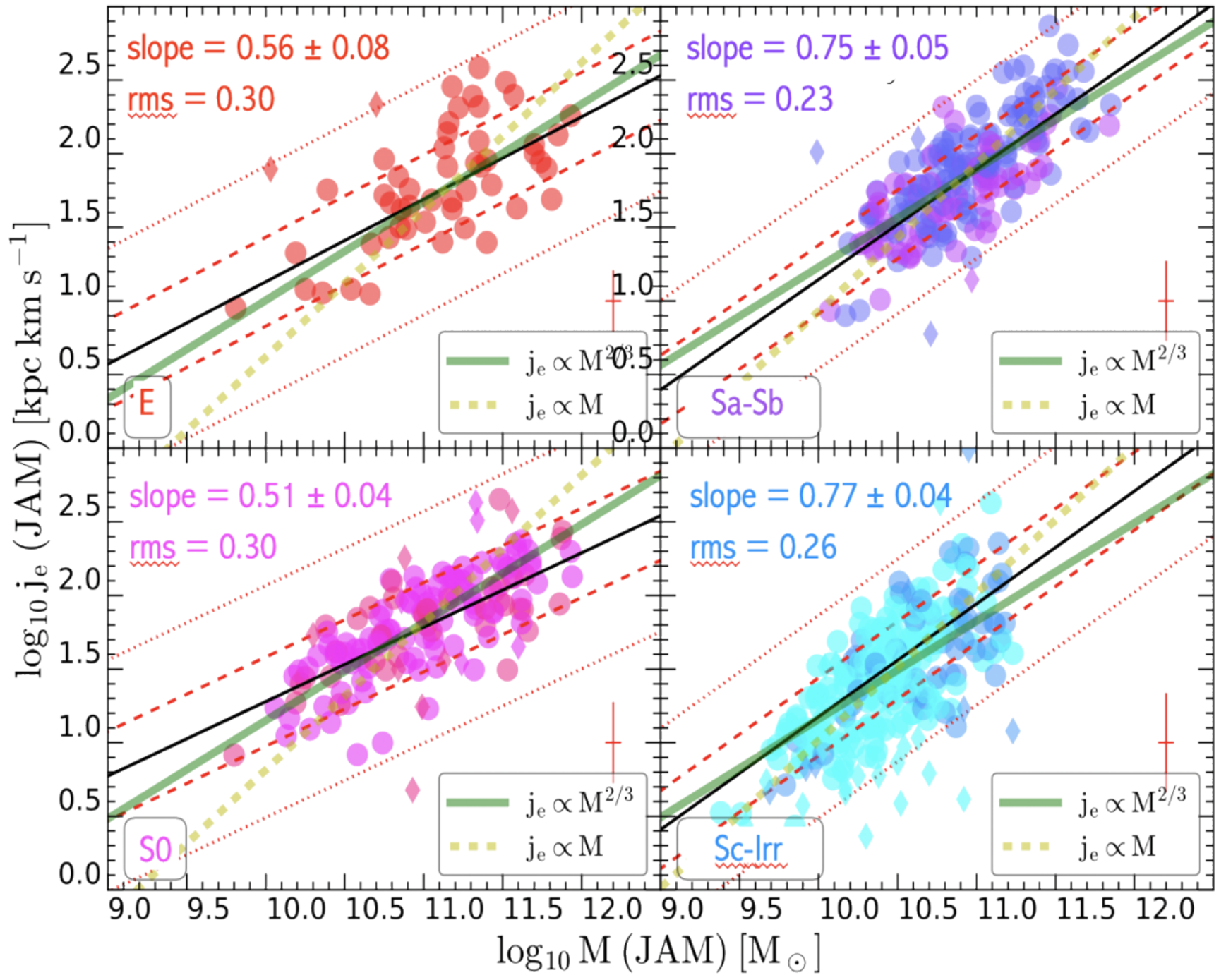}
  \caption{The relation between mass and specific angular momentum,
    both derived from Jeans anisotropic mass (JAM) models fitted to the 
    SAMI kinematic data, for elliptical (E), lenticular (S0), early-spiral
    (Sa-Sb) and late-spiral/irregular (Sc-Irr) galaxies. [D'Eugenio
    \etal, in prep.]} 
  \label{fig7}
\end{center}
\end{figure}

\section{Summary}

This is a golden age for studying galaxy angular momentum. Large
kinematic surveys using integral field spectrographs are vastly
increasing the amount and richness of the available information Sample
sizes are now beginning to allow studies of the dependence on multiple
simultaneous influences (mass/morphology/environment...)  The main
limitations remain instrumental trade-offs between spatial resolution
and radial coverage, and challenges in spatial resolution and surface
brightness at higher redshift.






\begin{thebibliography}{}
  
\bibitem[Bellstedt \etal\ (2018)]{Bellstedt+2018}{Bellstedt, S., Forbes,
    D.A., Romanowsky, A. J., \etal}, 
  2018, \textit{MNRAS}, 476, 4543
  [\href{https://doi.org/10.1093/mnras/sty456}{DOI:
    10.1093/mnras/sty456}]

\bibitem[Cortese \etal\ (2016)]{Cortese+2016}{Cortese, L., Fogarty,
    L.M.R., Bekki, K., \etal},
  2016, \textit{MNRAS}, 463, 170
  [\href{https://doi.org/10.1093/mnras/stw1891}{DOI:
    10.1093/mnras/stw1891}]

\bibitem[Falc\'on-Barroso (2016)]{Falcon-Barroso2016}{Falc\'on-Barroso,
    J.}, 2016, \textit{Astronomical Surveys and Big Data}, ASP Conf.\
  Series, 505, 133
  [\href{https://ui.adsabs.harvard.edu/\#abs/2016ASPC..505..133F}{ADS:
    2016ASPC..505..133F}]

\bibitem[Foster \etal\ (2018)]{Foster+2018}{Foster, C., van de Sande,
    J., Cortese, L., \etal},
    2018, \textit{MNRAS}, 480, 3105
  [\href{https://doi.org/10.1093/mnras/sty2059}{DOI:
    10.1093/mnras/sty2059}]

\bibitem[Graham \etal\ (2018)]{Graham+2018}{Graham, M.T., Cappellari,
    M., Li, H., \etal},
    2018, \textit{MNRAS}, 477, 4711
  [\href{https://doi.org/10.1093/mnras/sty504}{DOI:
    10.1093/mnras/sty504}]
  
\bibitem[Greene \etal\ (2018)]{Greene+2018}{Greene, J.E., Leauthaud, A.,
    Emsellem, E., \etal},
  2018, \textit{ApJ}, 852, 36
  [\href{https://doi.org/10.3847/1538-4357/aa9bde}{DOI:
    10.3847/1538-4357/aa9bde}]

\bibitem[Pulsoni \etal\ (2018)]{Pulsoni+2018}{Pulsoni, C., Gerhard, O.,
    Arnaboldi, \etal},
  2018, \textit{A\&A}, 618, 94
  [\href{https://doi.org/10.1051/0004-6361/201732473}{DOI:
    10.1051/0004-6361/201732473}]

\bibitem[Sanchez \etal\ (2015)]{Sanchez+2015}{S{\'a}nchez, Sebasti{\'a}n
    F., \& The CALIFA Collaboration}, 2015, \textit{Galaxies in 3D
    across the Universe}, IAU Symposium 309, pp85-92
  [\href{https://doi.org/10.1017/S1743921314009375}{DOI:
    10.1017/S1743921314009375}]

  \bibitem[van de Sande \etal\ (2018)]{vandeSande+2018}{van de Sande,
      J., Scott, N., Bland-Hawthorn, J., \etal},
    2018, \textit{Nature Astronomy}, 2, 483
  [\href{https://doi.org/10.1038/s41550-018-0436-x}{DOI:
    10.1038/s41550-018-0436-x}]

\end{thebibliography}
\end{document}